\documentclass[pdflatex,sn-mathphys-num]{sn-jnl}% Math and Physical Sciences Numbered Reference Style
\usepackage{graphicx}%
\usepackage{multirow}%
\usepackage{amsmath,amssymb,amsfonts}%
\usepackage{amsthm}%
\usepackage{mathrsfs}%
\usepackage[title]{appendix}%
\usepackage{xcolor}%
\usepackage{textcomp}%
\usepackage{manyfoot}%
\usepackage{booktabs}%
\usepackage{algorithm}%
\usepackage{algorithmicx}%
\usepackage{algpseudocode}%
\usepackage{listings}%
\theoremstyle{thmstyleone}%
\theoremstyle{thmstyletwo}%

\theoremstyle{thmstylethree}%

\begin{document}

\title[Article Title]{PepLLM: ESM-Guided Llama for Structured Protein--Peptide Binding Interface Analysis}

\author[1]{\fnm{Hao} \sur{Qian}}
% \email{qhonearth@sjtu.edu.cn}

\author*[1]{\fnm{Shikui} \sur{Tu}}\email{tushikui@sjtu.edu.cn}
\author[1]{\fnm{Lei} \sur{Xu}}
% \email{leixu@sjtu.edu.cn}

\affil[1]{ \orgdiv{Centre for Cognitive Machines and Computational Health (CMaCH)}, \orgdiv{School of Computer Science},
  \orgname{Shanghai Jiao Tong University}, \city{Shanghai}, \country{China}}


\abstract{
Protein--peptide interactions are central to cellular regulation and
peptide-based drug discovery, yet existing computational methods mainly focus on interaction classification, binding-site prediction, or peptide binder generation. These formulations provide limited insight into the physicochemical mechanisms that determine how a peptide binds to a protein. In this work, we introduce \textbf{PepLLM}, an instruction-tuned framework for structured protein--peptide interface understanding. Given protein--peptide sequences, PepLLM generates a machine-readable JSON annotation describing multiple interface properties, including peptide burial state, hydrogen-bond density, salt-bridge presence, hotspot residues, hydrophobicity, and electrostatic complementarity. To support this task, we construct a new protein--peptide interface dataset by integrating structural interface analysis, solvent-accessible surface area computation, hydrophobic burial estimation, electrostatic potential calculation, and redundancy-aware data splitting. PepLLM connects a pretrained ESM encoder with a LLaMA decoder through a nonlinear modality adapter. The adapted ESM residue embeddings are injected into the LLaMA prompt as continuous soft tokens via placeholder-token replacement, enabling the decoder to generate structured interface annotations under instruction tuning. By moving beyond single-label prediction toward multi-property and mechanism-aware generation, PepLLM establishes a new task and modeling paradigm for interpretable protein--peptide interface analysis.}

% \keywords{Protein--Peptide Interaction; Protein Language Model; Structured Interface Prediction}

\maketitle

\section{Introduction}

Protein--peptide interactions are fundamental to a wide range of biological
processes, including signal transduction, immune recognition, transcriptional
regulation, and enzyme inhibition. Short peptides often bind to protein surfaces
through transient and context-dependent interfaces, making them attractive
candidates for therapeutic design and molecular intervention. Compared with
protein--protein complexes, however, protein--peptide interactions are more
difficult to characterize computationally: peptides are highly flexible, their
binding poses can vary substantially across targets, and their binding affinity
is determined by a mixture of geometric, chemical, and electrostatic factors.
A useful computational model should therefore not only predict whether a peptide
binds to a protein, but also explain how the interface is formed.

Recent years have witnessed rapid progress in computational modeling of
protein--peptide interactions. Existing methods have addressed peptide-binding
site prediction, peptide--protein interaction classification, residue-level
interface prediction, and peptide binder generation
\citep{pepnn,pepcnn,tpeppro,pepmlm}. In parallel, structure-based geometric
learning models have improved the prediction of protein functional sites and
molecular interfaces from 3D structures \citep{scannet,pesto}. More recently,
protein language models and protein-oriented large language models have shown
strong potential for protein representation learning, protein function
prediction, and protein sequence generation
\citep{esm2,protllm,prollama,evollama}. Despite these advances, most existing
methods are still designed around a narrow prediction format, such as binary
interaction labels, binding residue scores, or generated binder sequences. They
do not directly produce a structured and interpretable description of the
physicochemical mechanisms underlying a protein--peptide interface.

This limitation is important because protein--peptide recognition is inherently
multi-faceted. A peptide may be deeply buried in a pocket or only weakly attached
to a protein surface. Its stability may be dominated by hydrogen bonds, salt
bridges, hydrophobic burial, electrostatic complementarity, or a small number of
hotspot residues. These properties are routinely considered by structural
biologists when interpreting docking poses or experimental complex structures,
but they are rarely unified into a single machine-learning task. Consequently,
there remains a gap between task-specific peptide--protein predictors and the
kind of mechanism-aware interface understanding required for downstream
applications such as peptide drug screening, docking result interpretation, and
interface engineering.

In this work, we introduce a new task, \emph{structured protein--peptide
interface understanding}. Given a protein--peptide complex represented through
sequence-level embeddings, the model is required to generate a structured JSON
object describing multiple interface properties, including peptide burial state,
hydrogen-bond density, salt-bridge presence, hotspot residues, hydrophobicity,
and electrostatic complementarity. This formulation differs from conventional
classification or residue-scoring tasks in two aspects. First, the output is
multi-property and mechanism-oriented rather than a single label. Second, the
prediction is expressed as a constrained, machine-readable object, making it
directly usable for automated evaluation, downstream filtering, and scientific
reasoning.

To support this task, we construct a new protein--peptide interface dataset from
complex structures. For each complex, we derive interface annotations by
combining structural interface analysis, solvent-accessible surface area
computation, hydrophobic burial estimation, electrostatic potential calculation,
and hotspot residue extraction. These heterogeneous structural signals are then
converted into unified JSON annotations. To reduce sequence-level leakage, we
perform redundancy-aware data splitting by clustering protein sequences and
assigning train, validation, and test sets at the cluster level. The resulting
dataset provides structured supervision for learning interpretable
protein--peptide interface properties rather than only predicting interaction
existence.

We further propose \textbf{PepLLM}, an instruction-tuned protein-language model
for structured protein--peptide interface understanding. PepLLM connects a
pretrained ESM encoder with a pretrained LLaMA decoder through a lightweight
nonlinear modality adapter. The ESM encoder produces residue-level
protein--peptide representations, which are projected into the LLaMA embedding
space and injected into the LLaMA prompt by replacing reserved placeholder-token
embeddings. The LLaMA decoder is then trained under teacher forcing to generate
the target JSON annotation. In this way, PepLLM treats protein--peptide sequence
representations as continuous soft tokens inside an instruction-following
language model, combining the biological knowledge of protein language models
with the structured generation ability of large language models.

Our main contributions are summarized as follows:
\begin{itemize}
\item We formulate \emph{structured protein--peptide interface understanding}
as a new task that requires unified prediction of multiple physicochemical
interface properties in a machine-readable format.

\item We construct a new protein--peptide interface dataset with annotations
derived from structural interface parsing, hydrophobic burial computation,
electrostatic complementarity estimation, hotspot extraction, and
redundancy-aware data splitting.

\item We propose \textbf{PepLLM}, an ESM--LLaMA instruction-tuning framework
that injects protein sequence embeddings into a large language model through
placeholder-based embedding replacement and generates structured interface
annotations.

\item We provide an ESM-only multi-head classification baseline to separate
the effect of protein sequence representation learning from the benefit of
instruction-based structured generation.

\end{itemize}

Together, our work moves protein--peptide modeling beyond binary interaction
prediction and binding-site localization toward interpretable, multi-property,
and mechanism-aware biomolecular reasoning.

\section{Method}
We propose \textbf{PepLLM}, an ESM--LLaMA instruction-tuning framework for
structured protein--peptide interface understanding. PepLLM bridges pretrained
protein sequence representations and large language model generation through a
lightweight modality adapter and a placeholder-based embedding injection
mechanism. The model first encodes the concatenated protein--peptide sequence
with ESM, projects the residue-level hidden states into the LLaMA embedding
space, and replaces reserved placeholder-token embeddings in the LLaMA prompt
with the projected protein embeddings. The LLaMA decoder is then trained to
generate a valid JSON object containing multiple physicochemical properties of
the protein--peptide interface.

% \paragraph{Contributions.}
% Our contributions are threefold. First, we formulate structured
% protein--peptide interface understanding as a new task that requires unified
% prediction of multiple physicochemical interface properties. Second, we build a
% new dataset of protein--peptide complexes with annotations derived from
% structural, surface, hydrophobic, and electrostatic analyses. Third, we propose
% \textbf{PepLLM}, an instruction-tuned ESM--LLaMA architecture that injects
% protein language model embeddings into a large language model and trains it to
% generate valid structured interface annotations. Together, these contributions
% move protein--peptide modeling beyond binary interaction prediction and toward
% interpretable, multi-property, language-model-based biomolecular reasoning.

\subsection{Dataset Preparation}

Each protein--peptide complex was represented by a structure file named in the form
\texttt{<pdbid>\_<pepChain>\_<protChain>.pdb}, from which the peptide chain and
protein chain were parsed directly. For each complex, the peptide and protein
sequences were extracted from the first structural model using Bio.PDB peptide
builders, and the final sequence input was constructed as
\[
x = x_{\mathrm{protein}} , \texttt{:} , x_{\mathrm{peptide}} .
\]
Samples with missing protein or peptide sequences were discarded. We further
removed complexes with protein length greater than 500 residues, peptide length
smaller than 2 residues, or a serialized target longer than 2000 tokens under the
\texttt{cl100k\_base} tokenizer. Only samples with valid electrostatic
complementarity statistics were retained.

\paragraph{Interface annotation from structural analysis.}
For each complex, PISA was first run in \texttt{--as-is} mode to identify
protein--peptide interfaces and to export detailed interface reports. The PISA
outputs were parsed to obtain buried solvent-accessible area, total
solvent-accessible area, hydrogen bonds, salt bridges, interfacing residues, and
key stabilizing or destabilizing residues. For an interface with buried areas
$B_1$ and $B_2$ on the two partners, the interface area was estimated as
\[
A_{\mathrm{int}} =
\begin{cases}
\dfrac{B_1+B_2}{2}, & B_1>0 \text{ and } B_2>0,\\
\max(B_1,B_2), & \text{otherwise}.
\end{cases}
\]
The peptide-like partner was assigned as the shorter partner, using residue count
when available and atom count as a fallback. The peptide burial ratio was defined as
\[
r_{\mathrm{burial}} =
\frac{B_{\mathrm{peptide}}}{A_{\mathrm{peptide}}},
\]
where $B_{\mathrm{peptide}}$ is the buried area of the peptide and $A_{\mathrm{peptide}}$ is its total solvent-accessible area. The peptide binding pose was discretized as
\[
\text{\texttt{burial\_state}} =
\begin{cases}
\text{\texttt{deep}}, & r_{\mathrm{burial}} \ge 0.60,\\
\text{\texttt{partial}}, & 0.30 \le r_{\mathrm{burial}} < 0.60,\\
\text{\texttt{surface}}, & r_{\mathrm{burial}} < 0.30.
\end{cases}
\]

Hydrogen bonds and salt bridges were normalized by the interface area:
\[
d_{\mathrm{HB}} = 100 \cdot \frac{n_{\mathrm{HB}}}{A_{\mathrm{int}}},
\qquad
d_{\mathrm{SB}} = 100 \cdot \frac{n_{\mathrm{SB}}}{A_{\mathrm{int}}}.
\]
The hydrogen-bond density label was set to \texttt{high} if
$d_{\mathrm{HB}}\ge 1.0$, \texttt{low} if $0<d_{\mathrm{HB}}<1.0$, and
\texttt{none} otherwise. Salt bridges were labeled as \texttt{present} if
$d_{\mathrm{SB}}>0$ and \texttt{absent} otherwise. Hotspot information was
collected from the PISA key-residue blocks by counting stabilizing and
destabilizing residues on both the protein and peptide sides.

\paragraph{Hydrophobic interface annotation.}
Hydrophobic burial was computed with FreeSASA. For each complex, the protein
chain, peptide chain, and full complex were evaluated separately to obtain total,
polar, and apolar solvent-accessible surface areas. The buried apolar area was
computed as
\[
\Delta A_{\mathrm{nonpolar}} =
\left(
A^{\mathrm{mono}}_{\mathrm{prot,apolar}} +
A^{\mathrm{mono}}_{\mathrm{pep,apolar}}
\right)
-
\left(
A^{\mathrm{complex}}_{\mathrm{prot,apolar}} +
A^{\mathrm{complex}}_{\mathrm{pep,apolar}}
\right).
\]
The corresponding hydrophobic free-energy contribution was estimated as
\[
\Delta G_{\mathrm{hydro}} =
-\gamma \Delta A_{\mathrm{nonpolar}},
\qquad
\gamma = 0.007~\mathrm{kcal\,mol^{-1}\,\AA^{-2}} .
\]
To obtain a categorical hydrophobicity label, we also computed the fraction of
buried area that was apolar:
\[
r_{\mathrm{hydro}} =
\frac{\Delta A_{\mathrm{nonpolar}}}
{
\left(
A^{\mathrm{mono}}_{\mathrm{prot,total}} +
A^{\mathrm{mono}}_{\mathrm{pep,total}}
\right)
-
\left(
A^{\mathrm{complex}}_{\mathrm{prot,total}} +
A^{\mathrm{complex}}_{\mathrm{pep,total}}
\right)
}.
\]
The hydrophobicity label was assigned as
\[
\text{\texttt{hydrophobicity}} =
\begin{cases}
\text{\texttt{hydrophobic}}, & r_{\mathrm{hydro}} \ge 0.60,\\
\text{\texttt{mixed}}, & 0.40 \le r_{\mathrm{hydro}} < 0.60,\\
\text{\texttt{polar}}, & r_{\mathrm{hydro}} < 0.40.
\end{cases}
\]

\paragraph{Electrostatic complementarity annotation.}
Electrostatic complementarity was computed from chain-specific Poisson--Boltzmann
electrostatic potentials. Each complex was split into its two chains, and each
chain was converted from PDB to PQR using \texttt{pdb2pqr} with the AMBER force
field at pH 7.0. APBS input boxes were generated automatically from the PQR
coordinates, and APBS was used to compute electrostatic potential maps for each
chain. Molecular surface vertices were generated by MSMS. Interface surface
points were defined as surface vertices within $4.0,\mathrm{\AA}$ of any atom
from the opposite chain.

The electrostatic potential of chain $B$ was sampled on the interface surface of
chain $A$, and vice versa. Nearest-neighbor surface-point pairs within
$3.5,\mathrm{\AA}$ were retained. Given paired potential values
$\phi_{B\rightarrow A}$ and $\phi_{A\rightarrow B}$, the Pearson correlation was
computed as
\[
\rho =
\mathrm{corr}(\phi_{B\rightarrow A}, \phi_{A\rightarrow B}),
\]
and electrostatic complementarity was defined as
\[
\mathrm{EC} = -\rho .
\]
Thus, negative correlation between opposing electrostatic potentials corresponds
to positive complementarity. We also computed the fraction of opposite-sign
potential pairs,
\[
f_{\mathrm{opp}} =
\frac{1}{N}\sum_{i=1}^{N}
\mathbf{1}\!\left\{
\phi_{B\rightarrow A}^{(i)}
\phi_{A\rightarrow B}^{(i)} < 0
\right\}.
\]
Cases with fewer than five valid interface points or fewer than five valid
paired points were treated as invalid. The final electrostatic label was assigned
from $\mathrm{EC}$ and $f_{\mathrm{opp}}$ as follows:
\[
C_{\mathrm{EC}} =
\begin{cases}
\text{\texttt{none}}, & f_{\mathrm{opp}} < 0.55,\quad \mathrm{EC} < 0.33,\\
\text{\texttt{weak}}, & f_{\mathrm{opp}} < 0.55,\quad 0.33 \le \mathrm{EC} < 0.50,\\
\text{\texttt{moderate}}, & f_{\mathrm{opp}} < 0.55,\quad \mathrm{EC} \ge 0.50,\\
\text{\texttt{weak}}, & 0.55 \le f_{\mathrm{opp}} \le 0.64,\quad \mathrm{EC} < 0.33,\\
\text{\texttt{moderate}}, & 0.55 \le f_{\mathrm{opp}} \le 0.64,\quad 0.33 \le \mathrm{EC} < 0.50,\\
\text{\texttt{strong}}, & 0.55 \le f_{\mathrm{opp}} \le 0.64,\quad \mathrm{EC} \ge 0.50,\\
\text{\texttt{moderate}}, & f_{\mathrm{opp}} > 0.64,\quad \mathrm{EC} < 0.33,\\
\text{\texttt{strong}}, & f_{\mathrm{opp}} > 0.64,\quad \mathrm{EC} \ge 0.33.
\end{cases}
\]

\paragraph{Target construction.}
For each retained complex, the structural annotations were merged into a single
JSON object. In the default setting, the model was trained to generate only the
coarse interface summary:
\begin{verbatim}
{
"burial_state": "surface|partial|deep",
"hbonds": {
"n_hbonds": 0,
"density": "none|low|high"
},
"salt_bridges": {
"n_salt_bridges": 0,
"presence": "absent|present"
},
"hotspots": {
"n_hotspots": 0
},
"hydrophobicity": "polar|mixed|hydrophobic",
"electrostatic_complementarity": "none|weak|moderate|strong"
}
\end{verbatim}
An optional full-task setting additionally included hydrogen-bond pairs,
salt-bridge pairs, and stabilizing or destabilizing hotspot residues on the
protein and peptide sides. Each final JSONL record contained the PDB identifier,
protein sequence, peptide sequence, and the above target annotations. When
multiple PISA interfaces were parsed for a complex, the first parsed interface
was used for the downstream JSONL record.

\paragraph{Redundancy-aware data split.}
To reduce sequence-level redundancy across splits, filtered protein sequences
were written to FASTA using the PDB identifier as the sequence name. The proteins
were clustered with MMseqs2 using
\texttt{--min-seq-id 0.3}, \texttt{-c 0.8}, \texttt{--cov-mode 1},
\texttt{-s 7.5}, and \texttt{--cluster-mode 2}. Splits were then assigned at the
cluster level rather than the individual-example level. Clusters with at least
1000 members were assigned to the training set. The remaining clusters were
bucketed by cluster size and randomly split with seed 42 into training,
validation, and test clusters using ratios 0.60, 0.23, and 0.17, respectively.
Buckets with fewer than 10 clusters were assigned to training to avoid unstable
small-bucket splits. The resulting split files were saved both as cluster-level
and flattened example-level JSON files.

\paragraph{Batch construction for ESM--LLaMA training.}
The dataset class loaded the JSONL records into memory and built an identifier
map from \texttt{pdbid} to example index. During collation, the protein and
peptide sequences were concatenated with a colon separator and tokenized by the
ESM tokenizer. If the concatenated sequence exceeded \texttt{max\_sequence\_length}, 
a random contiguous segment of length \texttt{max\_sequence\_length} was sampled. 
ESM inputs were truncated to \texttt{max\_sequence\_length} residues and 
right-padded to \texttt{max\_sequence\_length + 2} tokens, accounting for the 
beginning-of-sequence and end-of-sequence tokens.

For the language-model input, the target annotation was serialized as compact
JSON with sorted keys. The user prompt contained the text
\texttt{Sequence embeddings:} followed by a repeated reserved placeholder token.
The number of placeholder tokens was matched to the number of non-padding ESM
sequence tokens, so that the model could inject the ESM hidden states at the
corresponding positions in the LLaMA token sequence. The prompt was formatted
with the LLaMA chat template and left-padded. The target JSON was appended with
the LLaMA end-of-sequence token, tokenized without adding an extra BOS token, and
right-padded.

During training, the model input consisted of the concatenated prompt and target
tokens. The loss was applied only to the target JSON tokens: all prompt-token
labels and all padding-token labels were set to $-100$. Therefore, the model was
optimized to generate the structured interface annotation conditioned on the
protein--peptide sequence embeddings and the instruction prompt. During
inference, only the formatted prompt and ESM sequence tokens were provided, and
the serialized target was retained only for evaluation.

\subsection{Model Architecture and Training}

After constructing the protein--peptide interface annotation dataset, we trained
an instruction-following multimodal language model that maps protein--peptide
sequence representations to structured interface descriptions. The overall
framework consists of three modules: a pretrained ESM protein encoder, a
nonlinear modality adapter, and a pretrained LLaMA causal language model decoder.
The ESM encoder extracts residue-level contextual representations from the
protein--peptide sequence, the adapter projects these representations into the
LLaMA embedding space, and the LLaMA decoder generates the target interface
annotation in JSON format.

\paragraph{ESM encoder.}
For each input complex, the protein sequence and peptide sequence were
concatenated and tokenized by the ESM tokenizer. Given an input sequence
\[
s = (s_1, s_2, \ldots, s_L),
\]
the ESM encoder produces contextual hidden states
\[
H^{\mathrm{ESM}} = \mathrm{ESM}(s)
\in
\mathbb{R}^{B \times L \times d_{\mathrm{ESM}}},
\]
where $B$ is the batch size, $L$ is the padded sequence length, and
$d_{\mathrm{ESM}}$ is the hidden size of the ESM model. The ESM model was used
without an additional pooling layer, so residue-level hidden states were
preserved for downstream injection into the language model.

\paragraph{Modality adapter.}
Because the hidden dimension of ESM differs from the input embedding dimension
of LLaMA, we introduced a two-layer nonlinear adapter to align the two
representation spaces. For each residue hidden state $h_i^{\mathrm{ESM}}$, the
adapter computes
\[
u_i =
\mathrm{Dropout}
\left(
\mathrm{GELU}
\left(
W_1 h_i^{\mathrm{ESM}} + b_1
\right)
\right),
\]
\[
z_i =
\mathrm{Dropout}
\left(
\mathrm{GELU}
\left(
W_2 u_i + b_2
\right)
\right),
\]
followed by $\ell_2$ normalization:
\[
\tilde{z}_i =
\frac{z_i}{\left|z_i\right|_2}.
\]
The adapted representation is therefore
\[
Z =
(\tilde{z}_1, \tilde{z}*2, \ldots, \tilde{z}*L)
\in
\mathbb{R}^{B \times L \times d_{\mathrm{LLaMA}}},
\]
where $d_{\mathrm{LLaMA}}$ is the hidden size of the LLaMA decoder. In our
implementation, the adapter intermediate dimension was set to 2048.

\paragraph{Embedding-level injection into LLaMA.}
The adapted ESM representations were injected into LLaMA through a
placeholder-token replacement mechanism. Specifically, the user prompt contained
a reserved placeholder token repeated once for each non-padding ESM sequence
token. Let
\[
Y = (y_1, y_2, \ldots, y_T)
\]
denote the tokenized LLaMA input, including the system instruction, user prompt,
placeholder tokens, and, during training, the target JSON sequence. The standard
LLaMA embedding layer first maps these tokens to
\[
E =
\mathrm{Embed}_{\mathrm{LLaMA}}(Y)
\in
\mathbb{R}^{B \times T \times d_{\mathrm{LLaMA}}}.
\]
For positions corresponding to the reserved placeholder token, the original
token embeddings were replaced by the adapted ESM embeddings:
\[
\hat{E}_{b,t} =
\begin{cases}
Z_{b,k(t)}, & y_{b,t} = \langle\mathrm{placeholder}\rangle,\\
E_{b,t}, & \text{otherwise}.
\end{cases}
\]
where $k(t)$ indexes the corresponding valid ESM sequence position. The ESM
attention mask was used to ensure that only non-padding ESM hidden states were
inserted into the LLaMA context. This design allows the LLaMA decoder to attend
to protein--peptide sequence representations as continuous soft tokens, while
the surrounding instruction tokens remain standard textual embeddings.

\paragraph{Autoregressive decoder.}
The modified embedding sequence $\hat{E}$ was passed to the LLaMA causal decoder:
\[
p_\theta(Y_{\mathrm{out}} \mid s, Y_{\mathrm{prompt}}) =
\mathrm{LLaMA}(\hat{E}),
\]
where $Y_{\mathrm{prompt}}$ contains the system and user instructions, and
$Y_{\mathrm{out}}$ is the serialized JSON annotation. The output JSON contains
interface-level properties including peptide burial state, hydrogen-bond count
and density, salt-bridge count and presence, hotspot count, hydrophobicity, and
electrostatic complementarity. In the full-task setting, the output additionally
contains residue-pair and hotspot-residue details.

\paragraph{Parameter-efficient instruction tuning.}
The main ESM--LLaMA model was trained using parameter-efficient fine-tuning.
Pretrained ESM and LLaMA weights were loaded from their respective checkpoints,
and LoRA modules were inserted into the LLaMA decoder. LoRA was applied to the
self-attention projections
\[
q_{\mathrm{proj}};
k_{\mathrm{proj}};
v_{\mathrm{proj}};
o_{\mathrm{proj}}
\]
and the feed-forward projections
\[
\mathrm{gate}_{\mathrm{proj}};
\mathrm{up}_{\mathrm{proj}};
\mathrm{down}_{\mathrm{proj}}.
\]
The LoRA rank was set to $r=32$, the LoRA scaling factor was set to
$\alpha=2r$, and the LoRA dropout rate was set to 0.1. Unless otherwise stated,
the modality adapter was also trainable and saved together with the LoRA
parameters. This strategy updates only a small number of task-specific
parameters while retaining the pretrained sequence and language knowledge of ESM
and LLaMA.

\paragraph{Training objective.}
Training was performed with teacher forcing. For each example, the decoder input
contained the prompt followed by the target JSON sequence. The labels for prompt
tokens and padding tokens were set to $-100$, so the loss was applied only to
the JSON target tokens. The model was optimized using the standard causal
language modeling objective:
\[
\mathcal{L}_{\mathrm{LM}}
=
-\frac{1}{|\mathcal{T}|}
\sum_{t \in \mathcal{T}}
\log
p_\theta
\left(
y_t
\mid
y_{<t}, s
\right).
\]
where $\mathcal{T}$ denotes the set of target-token positions. This objective
encourages the model to generate valid structured interface annotations
conditioned on the instruction prompt and the ESM-derived protein--peptide
embeddings.

\paragraph{Distributed training protocol.}
Training was implemented with PyTorch DistributedDataParallel using the NCCL
backend on a single multi-GPU node. The dataset was partitioned across GPUs with
a distributed sampler; training batches were shuffled at each epoch, whereas
validation batches were not shuffled. Each dataloader used four CPU workers,
pinned memory, and dropped incomplete batches during training and validation.

Unless otherwise specified, the main ESM--LLaMA model was trained with the
following hyperparameters:
\begin{table}[t]
\centering
\begin{tabular}{ll}
\hline
ESM backbone & \texttt{facebook/esm2\_t36\_3B\_UR50D} \\
LLaMA backbone & \texttt{meta-llama/Meta-Llama-3.1-8B-Instruct} \\
Numerical precision & \texttt{bfloat16} \\
Batch size per GPU & 2 \\
Gradient accumulation steps & 8 \\
Learning rate & $2 \times 10^{-4}$ \\
Optimizer & Adam \\
Number of epochs & 100 \\
Learning-rate scheduler & StepLR \\
Scheduler decay factor & 0.8 \\
Random seed & 42 \\
LoRA rank & 32 \\
\hline
\end{tabular}
\end{table}
The loss was divided by the number of gradient-accumulation steps before
backpropagation, so each optimizer update corresponded to multiple micro-batches.
Gradient norm clipping was supported and disabled by default. After each epoch,
the learning-rate scheduler was stepped and the model was evaluated on the
validation split under teacher forcing. Training loss, validation loss,
learning rate, and gradient norm were recorded with TensorBoard. Adapter
checkpoints and optimizer/scheduler states were saved periodically.

\paragraph{Inference and generation.}
During inference, only the prompt and protein--peptide sequence were provided.
The ESM encoder first produced residue-level hidden states, the modality adapter
projected them into the LLaMA embedding space, and the projected embeddings
replaced the placeholder-token embeddings in the prompt. The resulting prompt
embeddings were passed to the LLaMA generation function to autoregressively
produce the output JSON. Before inference, the PEFT adapter was merged into the
base model for generation.

Generation was performed on the test split with a distributed sampler and
without shuffling. Unless otherwise specified, generation used greedy decoding
with one beam, no sampling, a maximum of 2000 newly generated tokens, temperature
1.0, top-$p$ equal to 1.0, and top-$k$ equal to 50. The end-of-sequence token ID
was set to 128009 and the padding token ID was set to 128002. Predictions and
ground-truth JSON strings were decoded with the LLaMA tokenizer and saved as
per-rank JSON files for downstream metric computation.

\paragraph{ESM-only classification baseline.}
To assess the contribution of the LLaMA decoder and instruction-based
generation, we also implemented an ESM-only classification baseline. This
baseline uses the same ESM encoder but removes the LLaMA decoder. The ESM
residue-level hidden states are first mean-pooled with the sequence attention
mask:
\[
h_{\mathrm{pool}} =
\frac{
\sum_{i=1}^{L} m_i h_i^{\mathrm{ESM}}
}{
\sum_{i=1}^{L} m_i
},
\]
where $m_i$ is the ESM attention mask. The pooled representation is then passed
to a multi-head adapter. Each prediction head is a two-layer nonlinear adapter
and outputs logits for one categorical interface property. The output dimensions
of the five heads were set to
\[
[3, 3, 2, 3, 4],
\]
corresponding to the class numbers of burial state, hydrogen-bond density,
salt-bridge presence, hydrophobicity, and electrostatic complementarity,
respectively.

For the ESM-only baseline, the training loss was the average cross-entropy over
all prediction heads:
\[
\mathcal{L}_{\mathrm{cls}} = 
\frac{1}{K}
\sum_{k=1}^{K}
\mathrm{CE}
\left(
\hat{y}^{(k)}, y^{(k)}
\right),
\]
where $K=5$. The baseline was trained using the same distributed training
framework. Unless otherwise specified, it used bfloat16 precision, Adam
optimization, learning rate $2 \times 10^{-4}$, 100 epochs, StepLR decay factor
0.8, gradient accumulation of 8 steps, and random seed 42. The batch size per
GPU was set to 4 for the ESM-only baseline. During inference, the predicted class
for each head was obtained by taking the argmax over the corresponding logits.

\section{Experiments}

\subsection{Experimental Setup}

We evaluate PepLLM on structured protein--peptide binding interface analysis.
Each input consists of a protein sequence and a peptide sequence, and each output
is a structured JSON object containing multiple interface properties. 

In the current experiments, we focus on a five-field classification setting covering
the following categorical attributes:
\begin{itemize}
    \item \texttt{burial\_state};
    \item \texttt{electrostatic\_complementarity};
    \item \texttt{hbonds.density};
    \item \texttt{hydrophobicity};
    \item \texttt{salt\_bridges.presence}.
\end{itemize}

These fields correspond to peptide burial state, electrostatic complementarity,
hydrogen-bond density, hydrophobic character, and salt-bridge presence,
respectively.

The dataset is constructed from protein--peptide complex structures and split
using protein-sequence clustering to reduce sequence-level leakage. After
filtering long sequences and clustering proteins with MMseqs2, the current
dataset contains approximately 32K training examples, 1.8K validation examples,
and 1.8K test examples. The protein sequence length is restricted to be less
than 500 residues, the peptide sequence length is restricted to be less than 50
residues, and the serialized JSON target is restricted to be shorter than 2000
tokens.

\begin{table}[t]
\centering
\caption{Dataset statistics used in the current experiments.}
\label{tab:dataset_stats}
\begin{tabular}{lr}
\toprule
Split & Number of examples \\
\midrule
Train & $\sim$32K \\
Validation & $\sim$1.8K \\
Test & $\sim$1.8K \\
\bottomrule
\end{tabular}
\end{table}

\paragraph{Models}
We compare two model families. The first is \textbf{PepLLM}, a generative model
that encodes the protein--peptide sequence pair with a pretrained protein
sequence encoder, projects the sequence embeddings through an MLP adapter, and
feeds the adapted embeddings into a general-purpose LLM decoder. The decoder is
trained to generate the target interface annotation as a JSON string.

The second model is an \textbf{ESM+adapter} baseline. This baseline removes the
LLM decoder and directly predicts the five categorical labels using a
multi-head adapter on top of the ESM sequence representation. This comparison is
designed to separate the contribution of structured LLM generation from the
protein sequence representation itself.

\paragraph{Metrics}
For the generative PepLLM model, we report the JSON pass rate and field-wise
accuracy. The JSON pass rate measures whether the generated string can be parsed
as a valid output object:
\[
\mathrm{PassRate}
=
\frac{1}{N}
\sum_{i=1}^{N}
\mathbb{I}
\left[
\mathrm{ValidJSON}(\hat{y}_i)
\right],
\]
For each categorical field $f$, we compute field accuracy:
\[
\mathrm{Acc}_f
=
\frac{1}{N}
\sum_{i=1}^{N}
\mathbb{I}
\left[
\hat{y}_{i,f} = y_{i,f}
\right].
\]
For count-valued fields, including the number of hydrogen bonds, hotspots, and
salt bridges, we additionally report mean absolute error (MAE):
\[
\mathrm{MAE}_f
=
\frac{1}{N}
\sum_{i=1}^{N}
\left|
\hat{y}_{i,f} - y_{i,f}
\right|.
\]

\subsection{Main Results}

Table~\ref{tab:pepllm_epoch_results} reports the performance of PepLLM after
one epoch and ten epochs of training. The model produces syntactically valid
JSON outputs with a very high pass rate throughout training. At Epoch 1, the
pass rate is 100.00\%, indicating that the model quickly learns the required
output schema. At Epoch 10, the pass rate remains high at 99.73\%.

Across the five categorical fields, PepLLM improves from an average accuracy of
49.14\% at Epoch 1 to 52.26\% at Epoch 10. 
The largest gains are observed for \texttt{hbonds.density}, which improves from 42.57\% to 48.11\%, and \texttt{burial\_state}, which improves from 60.93\% to 64.40\%. In contrast, \texttt{electrostatic\_complementarity} remains the most challenging attribute, suggesting that electrostatic interface labels may require more accurate supervision or richer structural context.

\begin{table}[t]
\centering
\caption{PepLLM results on the five-field classification setting. Accuracy is
reported in percentage. ``Avg.'' denotes the mean over the five categorical
fields.}
\label{tab:pepllm_epoch_results}
\begin{tabular}{lcc}
\toprule
Metric / Field & Epoch 1 & Epoch 10 \\
\midrule
JSON pass rate & 100.00 & 99.73 \\
\midrule
\texttt{burial\_state} & 60.93 & 64.40 \\
\texttt{electrostatic\_complementarity} & 34.08 & 35.18 \\
\texttt{hbonds.density} & 42.57 & 48.11 \\
\texttt{hydrophobicity} & 50.05 & 52.26 \\
\texttt{salt\_bridges.presence} & 58.07 & 61.36 \\
\midrule
Avg. categorical accuracy & 49.14 & 52.26 \\
\bottomrule
\end{tabular}
\end{table}

Table~\ref{tab:pepllm_mae_results} reports the MAE of count-valued JSON fields.
From Epoch 1 to Epoch 10, the MAE decreases for hydrogen-bond count and hotspot
count. The hotspot count MAE decreases from 4.15 to 3.69, suggesting that the
model benefits from additional training when predicting coarse interaction
intensity. The salt-bridge count MAE remains similar, increasing slightly from
1.56 to 1.61, which is consistent with the sparsity and imbalance of salt-bridge
events.

\begin{table}[t]
\centering
\caption{PepLLM results on count-valued JSON fields. Lower MAE is better.}
\label{tab:pepllm_mae_results}
\begin{tabular}{l c c}
\toprule
Count field & Epoch 1 & Epoch 10 \\
\midrule
\texttt{hbonds.n\_hbonds} & 3.09 & 2.92 \\
\texttt{hotspots.n\_hotspots} & 4.15 & 3.69 \\
\texttt{salt\_bridges.n\_salt\_bridges} & 1.56 & 1.61 \\
\midrule
Avg. MAE & 2.93 & 2.74 \\
\bottomrule
\end{tabular}
\end{table}

\subsection{Comparison with ESM-only Classification}

We further compare PepLLM with the ESM+adapter baseline on the same five
categorical fields. Table~\ref{tab:pepllm_vs_esm_adapter} summarizes the
results. The ESM+adapter baseline achieves an average accuracy of 54.21\%,
slightly higher than PepLLM's 52.26\%. However, the two models exhibit different
strengths. PepLLM performs better on \texttt{burial\_state},
\texttt{hbonds.density}, and \texttt{salt\_bridges.presence}, whereas the
ESM+adapter baseline performs better on \texttt{hydrophobicity} and
\texttt{electrostatic\_complementarity}.

This comparison suggests that simple discriminative heads can be competitive for
single-field classification, especially for global physicochemical attributes.
In contrast, PepLLM is designed for structured generation and can produce a
unified JSON output containing both categorical labels and count-valued
interaction information. Therefore, although the discriminative baseline is
strong for isolated classification fields, PepLLM provides a more general output
format for interpretable interface analysis.

\begin{table}[t]
\centering
\caption{Comparison between PepLLM and the ESM+adapter classification baseline
on the five categorical fields. Accuracy is reported in percentage.}
\label{tab:pepllm_vs_esm_adapter}
\begin{tabular}{lcc}
\toprule
Field & PepLLM & ESM+adapter \\
\midrule
\texttt{burial\_state} & \textbf{64.40} & 62.63 \\
\texttt{electrostatic\_complementarity} & 35.18 & \textbf{38.32} \\
\texttt{hbonds.density} & \textbf{48.11} & 48.09 \\
\texttt{hydrophobicity} & 52.26 & \textbf{62.10} \\
\texttt{salt\_bridges.presence} & \textbf{61.36} & 59.93 \\
\midrule
Avg. categorical accuracy & 52.26 & \textbf{54.21} \\
\bottomrule
\end{tabular}
\end{table}

% \subsection{Training Dynamics}

% The training curves show that PepLLM rapidly learns the output format and
% reduces training loss over optimization. However, the validation loss increases
% after several epochs, indicating possible overfitting under the current dataset
% scale and task formulation. This phenomenon is expected in the pilot setting:
% the model is large, the supervision is derived from automatically extracted
% structural annotations, and several labels are highly imbalanced. In particular,
% salt bridges are sparse and electrostatic complementarity is more difficult to
% infer from sequence representations alone.

% The ESM+adapter baseline exhibits a different trend. Its validation accuracy
% peaks in the early epochs and then decreases, suggesting that the discriminative
% baseline also overfits quickly. These observations motivate future experiments
% with stronger regularization, larger-scale cross-docking augmentation, improved
% label balancing, and structure-aware auxiliary inputs.

\subsection{Discussion}

The preliminary experiments lead to three observations. First, PepLLM can
reliably generate valid JSON outputs, showing that a general-purpose LLM can be
adapted to structured biomolecular interface analysis when conditioned on
protein--peptide sequence embeddings. Second, PepLLM improves over training on
most categorical fields and reduces count-field MAE, demonstrating that the
model learns nontrivial correlations between sequence-pair embeddings and
interface annotations. Third, the ESM+adapter baseline remains competitive on
some isolated categorical labels, especially hydrophobicity and electrostatic
complementarity, suggesting that future versions of PepLLM should combine
structured generation with stronger discriminative or auxiliary objectives.

Overall, these results validate the feasibility of the proposed task and
framework. While the current experiment is still preliminary, it demonstrates
that protein--peptide binding interface analysis can be formulated as a
structured generation problem rather than only as independent classification
tasks.

\section{Conclusion}

We presented \textbf{PepLLM}, a new dataset-driven instruction-tuning framework
for structured protein--peptide interface understanding. Unlike prior work that
mainly focuses on interaction classification, binding-site localization,
residue-level contact prediction, or peptide binder generation, PepLLM is
designed to generate a structured description of the underlying interface
mechanisms. By combining structural interface parsing, hydrophobic burial
estimation, electrostatic complementarity computation, and cluster-based data
splitting, we constructed a dataset that provides physicochemically grounded
supervision for protein--peptide interface analysis.

PepLLM connects pretrained protein sequence representations with large language
model generation. Through a lightweight modality adapter and placeholder-based
embedding injection, ESM-derived residue embeddings are inserted into the LLaMA
context as continuous soft tokens. The model is trained to generate valid JSON
annotations that summarize multiple interface properties in a unified format.
This framework demonstrates a practical way to repurpose large language models
for structured biomolecular prediction while retaining the biological knowledge
of protein language models.

More broadly, PepLLM suggests a new direction for computational protein science:
using language models not only to classify biological sequences or generate
candidate binders, but also to produce structured, interpretable, and
mechanism-aware descriptions of molecular interactions. Such a formulation can
support downstream applications including peptide drug screening, docking result
interpretation, interface engineering, and automated scientific hypothesis
generation. Future work may extend PepLLM by incorporating explicit 3D
structural encoders, jointly modeling multiple candidate binding poses, and
validating the generated interface annotations against experimental binding and
mutagenesis data.

\bibliography{sn-bibliography}% common bib file

\newpage

\begin{appendices}

\section{Cross-Docking Dataset Construction for Peptide--Protein Complexes}

\subsection{Overview}

Given a clustered peptide--protein complex dataset, the goal is to construct
cross-docked peptide--protein poses for downstream LLM training. Each complex is
identified as
\[
x = \texttt{pdbid\_peptideChain\_receptorChain}.
\]
For each cluster \(C_g\), each member can serve as a donor template and each
other member can serve as an acceptor target. A directed cross-docking pair is
defined as
\[
(d,a), \quad d,a \in C_g,\quad d \neq a,
\]
where \(d\) provides the template peptide and \(a\) provides the target receptor.

\subsection{Step 1: Standardization and Quality Control}

For each complex \(x\), the structure file is parsed from the input PDB/mmCIF
directory. The peptide chain and receptor chain are extracted according to the
identifier
\[
x = \texttt{pdbid\_peptideChain\_receptorChain}.
\]

Only the first model is used if the structure contains multiple models. For each
complex, three standardized PDB files are written:
\[
\texttt{complex\_raw.pdb},\quad
\texttt{peptide.pdb},\quad
\texttt{receptor.pdb}.
\]

The following quality-control features are computed for both peptide and
receptor chains:
\[
\ell = \text{number of amino-acid residues},
\]
\[
n_{\mathrm{nonstd}} = \text{number of non-standard residues},
\]
\[
n_{\mathrm{missBB}} = \#\{r: \{N,CA,C\} \nsubseteq \mathrm{Atoms}(r)\},
\]
\[
n_{\mathrm{break}} =
\#\{(r_i,r_{i+1}): \|CA_i - CA_{i+1}\|_2 > 4.5\ \text{\AA}\}.
\]

Each complex is assigned a status:
\[
\mathrm{status}(x) \in \{\texttt{ok},\texttt{repairable},\texttt{failed}\}.
\]
A complex is marked as failed if the structure is missing, cannot be parsed, has
missing required chains, or has an empty peptide or receptor chain. It is marked
as repairable if backbone atoms are missing but the chains can still be exported.

The outputs of this step are:
\[
\texttt{complex\_table.csv},\quad
\texttt{pair\_table.csv},\quad
\texttt{cluster\_table.csv},\quad
\texttt{qc\_report.csv}.
\]

For each cluster \(C_g\), the initial directed pair set is
\[
\mathcal{P}_g =
\{(d,a): d,a \in C_g,\ d \neq a\},
\]
with size
\[
|\mathcal{P}_g| = |C_g|(|C_g|-1).
\]

\subsection{Step 1.5: Pair Pre-screening and Top-\(K\) Re-ranking}

Because the full directed pair table can be close to \(O(n^2)\), a lightweight
two-stage ranking procedure is used before docking.

For each standardized complex \(x\), peptide and receptor sequences are
extracted. The receptor binding site is defined by heavy-atom contacts to the
native peptide:
\[
S_x =
\left\{
i \in R_x:
\min_{j \in P_x}
\min_{\alpha,\beta}
\left\|
r_{i,\alpha} - p_{j,\beta}
\right\|_2
\leq 6.0\ \text{\AA}
\right\}.
\]
The contact site is then expanded by a sequence padding window:
\[
S_x^{+} = \{i-k,\ldots,i+k: i \in S_x,\ k=1\}.
\]

A donor-quality score is computed from the Step 1 QC features:
\[
\begin{aligned}
q_d
=
\mathrm{clip}_{[0,1]}
\Big(&
1
- \Delta_{\mathrm{status}}
- 0.03\min(5,n_{\mathrm{pepMissBB}})
- 0.02\min(10,n_{\mathrm{recMissBB}}) \\
&- 0.01\min(10,n_{\mathrm{pepNonstd}})
- 0.005\min(10,n_{\mathrm{recNonstd}})
- 0.05\min(5,n_{\mathrm{pepBreak}}) \\
&- 0.03\min(5,n_{\mathrm{recBreak}})
\Big).
\end{aligned}
\]
where
\[
\Delta_{\mathrm{status}} =
\begin{cases}
0, & \texttt{ok},\\
0.15, & \texttt{repairable},\\
1.0, & \texttt{failed}.
\end{cases}
\]

For each acceptor \(a\), all possible donors \(d\) in the same cluster are
cheaply pre-scored. Let \(v^P_x\) be the amino-acid composition vector of the
peptide sequence, and \(v^S_x\) be the composition vector of the receptor
binding-site sequence. Define
\[
\rho^P_{d,a}
=
\frac{\min(\ell^P_d,\ell^P_a)}
     {\max(\ell^P_d,\ell^P_a)},
\qquad
\rho^S_{d,a}
=
\frac{\min(\ell^S_d,\ell^S_a)}
     {\max(\ell^S_d,\ell^S_a)}.
\]

A candidate pair is valid only if
\[
d \neq a,
\]
\[
\rho^P_{d,a} \geq 0.60,
\]
\[
|\ell^P_d - \ell^P_a| \leq 15,
\]
\[
\rho^S_{d,a} \geq 0.50,
\]
and both peptide and site lengths are non-zero.

The cheap pre-score is
\[
s^{\mathrm{pre}}_{d,a}
=
0.35\cos(v^S_d,v^S_a)
+
0.25\cos(v^P_d,v^P_a)
+
0.15\rho^P_{d,a}
+
0.15\rho^S_{d,a}
+
0.10q_d.
\]

For each acceptor, only the top \(M\) donors by \(s^{\mathrm{pre}}\) are kept,
where the default is
\[
M = 100.
\]

These candidates are then re-ranked using exact global sequence identities:
\[
I^S_{d,a} = \mathrm{SeqId}(S_d,S_a),
\qquad
I^P_{d,a} = \mathrm{SeqId}(P_d,P_a).
\]

The final ranking score is
\[
s^{\mathrm{final}}_{d,a}
=
0.45s^{\mathrm{pre}}_{d,a}
+
0.30I^S_{d,a}
+
0.20I^P_{d,a}
+
0.05q_d.
\]

For each acceptor, the final top \(K\) donors are selected:
\[
\mathcal{T}_a =
\operatorname{TopK}_{d}
\left(
s^{\mathrm{final}}_{d,a}
\right),
\qquad K=25.
\]

The outputs are:
\[
\texttt{complex\_feature\_table.csv},
\quad
\texttt{pair\_table\_ranked.csv},
\quad
\texttt{pair\_table\_topk.csv}.
\]

\subsection{Step 2: Local Binding-Site Alignment and Initial Model Construction}

For each selected donor--acceptor pair \((d,a)\), the standardized receptor and
peptide structures are loaded.

The donor and acceptor native binding sites are recomputed using the same
heavy-atom contact rule:
\[
S_d =
\left\{
i \in R_d:
\min_{j \in P_d}
\min_{\alpha,\beta}
\|r^d_{i,\alpha} - p^d_{j,\beta}\|_2
\leq 6.0\ \text{\AA}
\right\},
\]
\[
S_a =
\left\{
i \in R_a:
\min_{j \in P_a}
\min_{\alpha,\beta}
\|r^a_{i,\alpha} - p^a_{j,\beta}\|_2
\leq 6.0\ \text{\AA}
\right\}.
\]

The donor and acceptor receptor sequences are globally aligned:
\[
A(R_d,R_a) \rightarrow \pi_{d \rightarrow a},
\]
where \(\pi_{d \rightarrow a}\) maps donor receptor residue indices to acceptor
receptor residue indices.

Mapped site pairs are constructed as
\[
\mathcal{M}_{d,a}
=
\{(i,\pi_{d\rightarrow a}(i)):
i \in S_d,\ \pi_{d\rightarrow a}(i) \in S_a,\ CA_i,CA_{\pi(i)}\ \text{exist}\}.
\]

If
\[
|\mathcal{M}_{d,a}| < 4,
\]
the pipeline attempts a fallback using all full-chain mapped \(C_\alpha\) pairs.
If the fallback also contains too few pairs, the cross-docking pair is rejected.

A rigid-body superposition is computed by minimizing
\[
(R^\ast,t^\ast)
=
\arg\min_{R,t}
\sum_{(i,j)\in \mathcal{M}_{d,a}}
\left\|
R\,CA^d_i + t - CA^a_j
\right\|_2^2.
\]

The donor peptide is transferred into the acceptor receptor coordinate system:
\[
P_{d\rightarrow a}
=
R^\ast P_d + t^\ast.
\]

The initial cross-docked model is then defined as
\[
M^{0}_{d,a}
=
R_a \cup P_{d\rightarrow a}.
\]

Importantly, the acceptor native peptide is saved only as a reference and is not
used for model construction:
\[
P_a^{\mathrm{native}} \notin M^{0}_{d,a}.
\]

For each pair, the following metadata are recorded:
\[
\mathrm{local\_RMSD},\quad
\mathrm{receptor\_seq\_identity},\quad
\mathrm{peptide\_seq\_identity},\quad
n_{\mathrm{clashResidues}},\quad
n_{\mathrm{clashAtomPairs}}.
\]

The main output is:
\[
\texttt{pair\_summary.csv},
\]
and, when model writing is enabled,
\[
\texttt{initial\_model.pdb}.
\]

\subsection{Step 3: ADCP Local-Docking Job Selection}

Only successful Step 2 pairs with existing initial models are considered for
ADCP local docking. A pair is accepted if it satisfies:
\[
\ell^P_d \leq 50,
\]
\[
|\mathcal{M}_{d,a}| \geq 4,
\]
\[
\mathrm{local\_RMSD}_{d,a} \leq 8.0\ \text{\AA},
\]
\[
n_{\mathrm{clashResidues}} \leq 12.
\]

If an extrapolation ratio is available, the additional filter is
\[
\mathrm{extrap\_ratio}_{d,a} \leq 0.50.
\]

Optionally, the pipeline can require
\[
\mathrm{alignment\_mode} = \texttt{local\_site},
\]
and can reject trivial native-like cases where
\[
\mathrm{SeqId}(R_d,R_a)=1
\quad\text{and}\quad
\mathrm{SeqId}(P_d,P_a)=1.
\]

Accepted jobs are written to
\[
\texttt{jobs.csv}.
\]

\subsection{Step 4: Preparing PDB Inputs for HPC Docking}

For each accepted job, the initial model is copied into a flat docking-input
directory:
\[
\texttt{step4/\{pair\_id\}.pdb}.
\]

For large-scale HPC transfer, the PDB files can be split into file lists,
packed into tar chunks, transferred to the remote cluster, and extracted into
the remote Step 4 directory:
\[
\texttt{step4\_tar\_chunks/chunk\_*.tar}
\rightarrow
\texttt{remote/step4/}.
\]

\subsection{Step 5: ADCP Local Docking}

For each input complex \(M^{0}_{d,a}\), the ADCP driver performs the following
operations.

First, the complex is split into receptor and peptide PDB files:
\[
M^{0}_{d,a}
\rightarrow
\texttt{receptor.pdb},\quad
\texttt{peptide.pdb}.
\]

The peptide sequence is extracted from the peptide PDB:
\[
P_{d\rightarrow a}
\rightarrow
\mathrm{seq}(P_d).
\]

Both receptor and peptide are protonated:
\[
\texttt{receptor.pdb}
\xrightarrow{\texttt{reduce}}
\texttt{receptorH.pdb},
\]
\[
\texttt{peptide.pdb}
\xrightarrow{\texttt{reduce}}
\texttt{peptideH.pdb}.
\]

They are then converted to PDBQT:
\[
\texttt{receptorH.pdb}
\xrightarrow{\texttt{prepare\_receptor}}
\texttt{receptorH.pdbqt},
\]
\[
\texttt{peptideH.pdb}
\xrightarrow{\texttt{prepare\_ligand}}
\texttt{peptideH.pdbqt}.
\]

A local docking target is built around the current peptide pose:
\[
\texttt{agfr}
\left(
\texttt{receptorH.pdbqt},
\texttt{peptideH.pdbqt},
\mathrm{padding}=4.0\ \text{\AA}
\right)
\rightarrow
\texttt{job.trg}.
\]

ADCP is then run as
\[
\texttt{adcp}
\left(
\texttt{-t job.trg},
\texttt{-s peptide\_sequence},
\texttt{-N 20},
\texttt{-n 100000},
\texttt{-ref peptideH.pdb}
\right).
\]

The output is a ranked set of docked peptide poses:
\[
\texttt{job\_cdock\_ranked\_*.pdb}.
\]

On SLURM, the job table is distributed across workers by row index:
\[
\mathrm{worker}(i) = i \bmod N_{\mathrm{tasks}}.
\]
The provided batch script launches
\[
5\ \text{nodes} \times 10\ \text{tasks per node} = 50\ \text{workers},
\]
with 6 CPU cores per worker.

\subsection{Step 6: Post-processing and Final Complex Reconstruction}

After ADCP, each docked peptide pose is merged back with the corresponding
protein receptor.

The protein chain is extracted from the Step 4 initial complex, and the docked
peptide chain is extracted from the ADCP output. The final chain identifiers are
normalized as
\[
\text{peptide chain} = A,
\qquad
\text{protein chain} = B.
\]

The final reconstructed complex is
\[
M^{\mathrm{dock}}_{d,a,k}
=
P^{\mathrm{dock}}_{d,a,k}
\cup
R_a,
\]
where \(k\) indexes the ADCP ranked pose.

The final outputs are stored in
\[
\texttt{step6/\{pair\_id\}/\{ranked\_pose\}.pdb}.
\]

\subsection{LLM Training Record Construction}

Each final training example can be represented as a structured record:
\[
z =
\left(
\mathrm{input}_{d,a},
\mathrm{target}_{d,a,k},
\mathrm{metadata}_{d,a,k}
\right).
\]

The input can include:
\[
\mathrm{input}_{d,a}
=
\{
\mathrm{seq}(R_a),
\mathrm{seq}(P_d),
M^{0}_{d,a},
\mathrm{cluster\_id},
\mathrm{donor\_id},
\mathrm{acceptor\_id},
s^{\mathrm{final}}_{d,a}
\}.
\]

The target can be the final docked complex or the docked peptide coordinates:
\[
\mathrm{target}_{d,a,k}
=
M^{\mathrm{dock}}_{d,a,k}
\quad
\text{or}
\quad
P^{\mathrm{dock}}_{d,a,k}.
\]

The metadata should include:
\[
\{
\mathrm{local\_RMSD},
\mathrm{site\_pair\_n},
\mathrm{peptide\_length},
\mathrm{clash\_counts},
\mathrm{ADCP\_rank},
\mathrm{source\_cluster}
\}.
\]

To avoid information leakage, train/validation/test splitting should be
performed at the cluster level:
\[
C_g \cap C_h = \emptyset,
\qquad
g \in \mathcal{G}_{\mathrm{train}},
\quad
h \in \mathcal{G}_{\mathrm{test}}.
\]
All cross-docking pairs derived from the same cluster should remain in the same
dataset split.

Overall, the pipeline generated approximately $30,000$
directed peptide--protein cross-docking pairs, providing a large-scale training
corpus for LLM-based peptide--protein binding and docking modeling.

\subsection{Implementation Notes}

\begin{enumerate}
    \item Step 1.5 writes both \texttt{pair\_table\_ranked.csv} and
    \texttt{pair\_table\_topk.csv}. For large-scale docking, the top-\(K\)
    file should normally be used as the Step 2 input.

    \item Step 2 must be executed with model writing enabled; otherwise
    \texttt{initial\_model.pdb} will not exist and Step 3 will reject the job.

    \item The ADCP job preparation step uses local RMSD, site-pair count,
    peptide length, and clash count as heuristic filters. These thresholds
    should be tuned for the desired balance between throughput and quality.

    \item The final post-processing step assumes that the docked peptide chain
    in the ADCP output is chain \(A\), and rewrites the final complex as
    peptide chain \(A\) and protein chain \(B\).
\end{enumerate}

\end{appendices}
\end{document}